\documentclass[aps,pra,twocolumn,twoside,superscriptaddress]{revtex4}

\usepackage{bbm}
\usepackage{amsmath}
\usepackage{color}
\usepackage{graphicx,epsfig}

\newcommand{\tr}{{\textrm{tr}}}

\begin{document}

\title{How much of one-way computation is \textit{just} thermodynamics?}

\author{Janet Anders\footnote{email: janet@qipc.org}}
\affiliation{	Centre for Quantum Technologies, National University of Singapore, 3 Science Drive 2, Singapore 117543.}
\affiliation{Department of Physics and Astronomy, University College London, London WC1E 6BT, United Kingdom.}

\author{Michal Hajdu\v{s}ek}
\affiliation{The School of Physics and Astronomy, University of Leeds, Leeds LS2 9JT, United Kingdom.}

\author{Damian Markham}
\affiliation{Department of Physics, Graduate School of Science, University of Tokyo, Tokyo 113-0033, Japan.}
\affiliation{Universit\'e Paris 7, 175 Rue du Chevaleret, 75013 Paris, France.}

\author{Vlatko Vedral}
\affiliation{The School of Physics and Astronomy, University of Leeds, Leeds LS2 9JT, United Kingdom.}
\affiliation{	Centre for Quantum Technologies, National University of Singapore, 3 Science Drive 2, Singapore 117543.}

\date{18 June 2008}

\begin{abstract}
In this paper we argue that one-way quantum computation can be seen as a form of phase transition with the available information about the solution of the computation being the order parameter. We draw a number of striking analogies between standard thermodynamical quantities such as energy, temperature, work, and corresponding computational quantities such as the amount of entanglement, time, potential capacity for computation, respectively. Aside from being intuitively pleasing, this picture allows us to make novel conjectures, such as an estimate of the necessary critical time to finish a computation and a proposal of suitable architectures for universal one-way computation in 1D.


\medskip

\end{abstract}

\maketitle

\section{Motivation}

In a thought-provoking paper \cite{Toffoli}, Toffoli argued that physics is computation. He achieves this by mapping the Lagrangian `action' in physics, which counts in how many different ways a system can evolve dynamically, to Shannon's entropy in information theory, which quantifies the lack of information one has about the state of a system. So, ``action is to dynamics exactly what entropy to statics'' \cite{Toffoli}. In this paper  we will invert Toffoli's logic and map the process of a one-way quantum computation \cite{RaussendorfBriegel} to a physical description in the language of thermodynamics. Ours is a conceptual and heuristic paper whose main aim is to put one-way computation in a much broader physical framework. An exciting consequence of this idea will be that one-way quantum computation can be viewed as a phase transition from a disordered phase, with diverse possible outcomes of the computation, to an ordered phase, where the unique solution to the computation is reached. 


Adopting this view has a number of merits. First of all, some already known results in one-way quantum computation become transparent when formulated with the view that the number of ways a computation can be implemented determines computational capacity. For example, the fact that 1D one-way cluster computer is not universal as well as the fact that the 2D cluster computer can be. Secondly, it allows us to explain some desirable properties of the one-way computer design without having to use any heavy mathematical machinery. Here we will address the issue of the efficiency of one-way computers as a function of their inter-connectivity. 


One-way quantum computation relies on the initial, highly entangled state of many qubits, for instance the cluster state, which forms the resource for the computation. An algorithm can be implemented by measuring a fraction of the state's qubits in a specific and controlled way thereby driving the rest of the `computer', i.e. the state of all qubits,  to produce the solution to the algorithm. Since the computation is driven by measurements and not by unitaries, the cluster state computer is not reversible, which is why it is referred to as one-way computer \footnote{We will use one-way computer and cluster state computer as synonymous. Strictly speaking, the former is a broader concept involving general graph states, with the cluster state being a special case of them. This subtle distinction will be irrelevant for our discussion.}.
The one-way computer can only work by way of the entanglement that correlates qubits inside the cluster state and `spreads' information about the performed measurements over all qubits in the computer.  Entanglement is thus the true resource of the computation which is used up to produce the `solution', i.e. bits of information. It is, in this respect, analogous to solid state systems that trade energy for reduced entropy when they undergo a phase transition.


Moreover, the one-way model of computation is fully equivalent in power and capacity to the standard model of a quantum Turing machine \cite{RaussendorfBriegel}. Understanding the key ingredients and the dynamics of one-way computing by formulating it as a critical physical process will shed light on what the power of quantum computing is, that goes beyond the realm of classical computing.
Furthermore, the measurement-based aspect of one-way computing offers practical advantages over the conventional gate model in a number of experimental settings making it the model of choice for quantum computing. A recent experiment has already demonstrated the first implementation of basic gates and the Grover search algorithm using entangled photons as cluster states \cite{Walther}. Finally, discussing the one-way quantum computer as a thermal system will help identifying new candidates for realistic experimental implementations.


The paper is structured as follows. We summarise the basics of one-way computing with the cluster state as the resource in Section \ref{sec:one-way}. In Section \ref{sec:Peierls} we briefly explain the idea behind the concept of phase transitions and review Peierls' argument of why there is no phase transition in the 1D Ising model and why there can be in higher dimension.  We present our main ideas in Section \ref{sec:analogy} where we propose a mapping of relevant thermodynamical quantities to central computational characteristics, in particular the free energy to the computational potential. In Section \ref{sec:conclusions} we derive first results from the established heuristic analogy and give an outlook on further questions that could be addressed using the same link.

\section{One-way cluster computing} \label{sec:one-way}

A possible initial state of the one-way computer, the cluster state, is prepared in the following way. First take an array of $N$ qubits arranged in a $d$-dimensional square lattice. Then prepare each of these qubits in an equal superposition state with respect to the $z$-basis, $| 0 \rangle$ and $|1 \rangle$, as 
$| + \rangle = \frac{1}{\sqrt{2}}(| 0 \rangle + | 1 \rangle)$.
Finally, apply a two-qubit C-phase gate, $CZ$, to every pair of nearest neighbour qubits in the lattice, changing, for instance, $|++ \rangle$ into the entangled state
\begin{equation}
	CZ | + \rangle | + \rangle = 
	\frac{1}{\sqrt{2}}(| 0 \rangle | + \rangle + | 1 \rangle | - \rangle).
\end{equation}
The full two-qubit C-phase operator can be written in matrix form with respect to the basis $|00\rangle, |01\rangle, |10\rangle, |11\rangle$ as
\begin{equation} \label{eq:cz}
CZ=\left(
     \begin{array}{cccc}
       1 & 0 & 0 & 0 \\
       0 & 1 & 0 & 0 \\
       0 & 0 & 1 & 0 \\
       0 & 0 & 0 & -1 \\
     \end{array}
   \right).
\end{equation}
The resulting fully C-phased, highly entangled $N$-qubit state is the \emph{cluster state} which can be depicted schematically as below for $N=16$ in the two-dimensional case. 
\begin{center}	
	\includegraphics[width=2cm]{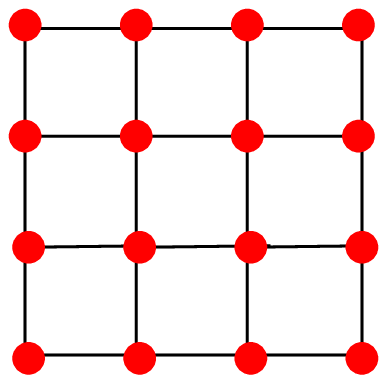} 
\end{center}
The red dots represent the qubits and the lines connecting them signify that they have been C-phased. Note, that the order of applying the C-phase gates does not matter since the C-phase operators commute. Experimentally such an operation could be achieved by letting an Ising-type Hamiltonian act on the whole lattice for an appropriately chosen time.


The amount of entanglement in the cluster state is directly proportional to the number of applied C-phase gates\footnote{Note, that although the whole cluster state is entangled, no two qubits are entangled by themselves (i.e. when all other qubits are traced out, the two remaining  qubits are in a separable configuration). Instead,  every applied C-phase gate adds roughly one unit of entanglement to the cluster which is uniformly shared across the cluster.}. A suitable measure of entanglement is the \emph{relative entropy of entanglement} \cite{RelEntEnt}. The relative entropy of entanglement of a state $\rho$ is defined as $E_{R}(\rho)= \min_{\sigma \in D} \, S(\rho\|\sigma)$, where $S(\rho\|\sigma)= \tr[ \rho \, \textrm{log} \, \rho - \rho \, \textrm{log} \, \sigma]$ is the quantum relative entropy and the minimisation is taken over all disentangled mixed states $\sigma$. To illustrate that the relative entropy of entanglement is a useful measure in the context of cluster states consider the box cluster state,  which has the mathematical form
\begin{eqnarray}\label{eq:boxcluster}
	|\textrm{box-cluster}\rangle &=& |0\rangle |+\rangle |0\rangle |+\rangle 
	+ |0\rangle |-\rangle |1\rangle |-\rangle \nonumber \\
 	&+&  |1\rangle |-\rangle |0\rangle |-\rangle 
	+ |1\rangle |+\rangle|1\rangle |+\rangle.
\end{eqnarray}
Pictorially the box cluster can be represented as a square with one of the four qubits at each vertex.
\begin{center}
	\includegraphics[width=1cm]{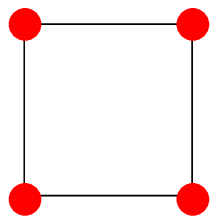}
\end{center}
All cluster states of any size are pure states and their relative entropy reduces to $S(\rho\|\sigma)= -\tr[\rho ~ \textrm{log} ~\sigma]$. The relative entropy  of entanglement then requires to find the disentangled mixed state $\sigma$ that minimises the relative entropy. It can be shown that this minimum is reached for $\sigma=\textrm{diag}\{\frac{1}{16},\ldots,\frac{1}{16}\}$ so that the relative entropy of entanglement of the box cluster becomes $E_R(|\textrm{box-cluster}\rangle) = 4$. This coincides with the number of applied C-phase gates. The same holds for general cluster states where the entanglement is always proportional to the number of applied C-phase gates \cite{Markham}.


To see the mechanism of one-way computing at work, consider the simplest case of a two qubit cluster state computer (see \cite{RaussendorfBriegel} as the general reference). Instead of being prepared in $|+\rangle$, the first qubit is the input and can be any state $|\psi\rangle = \alpha |0 \rangle + \beta |1 \rangle$, with $\alpha$ and $\beta$ arbitrary complex numbers with $|\alpha|^2 + |\beta|^2 = 1$. The second qubit is initially prepared in the state $|+ \rangle$ and will serve as the output state of the computation. The two qubits are C-phased and become entangled in the total state of both qubits, $\alpha|0\rangle|+\rangle + \beta|1\rangle|-\rangle$. 
As a computational step we now apply a Hadamard gate, 
$H =  {1 \over \sqrt{2}}
	\left( \begin{array}{cc} 
		1 & \phantom{-} 1 \\ 1 & -1 \end{array} \right),$ 
to the first qubit which leads to new total state $\alpha|+\rangle|+\rangle + \beta|-\rangle|-\rangle$. Rewriting this state we find that it can be cast in the form $\frac{1}{\sqrt{2}}[|0\rangle\otimes H|\psi\rangle + |1\rangle\otimes \sigma_x \, H \, |\psi\rangle]$, with $\sigma_x$ being the usual Pauli matrix. This implies that a measurement of the first qubit in the computational $z$-basis will in any event have the effect of teleporting the input state to the second qubit - up to a Pauli transformation. Indeed, the state of the second qubit after measuring the first qubit can be written more compactly as $\sigma_x^{m} \, H \, |\psi\rangle$ where we have introduced the new variable $m\in\{0,1\}$. The variable $m$ corresponds to the two possible outcomes of our measurement in the computational basis. After measuring the first qubit and finding either one of the two values, we have to correct and apply the inverse $H \, \sigma_x^{m}$ to obtain the fully teleported state $|\psi \rangle$. 


This simple example can be generalised to implement any rotation $R_{z}(\phi)$ of the arbitrary input qubit where the angle of the rotation is varied by using the basis $\{\frac{1}{\sqrt{2}}(|0\rangle\pm e^{i\phi}|1\rangle)\}$ for measurement. This will leave the second qubit in the desired rotated state,  $\sigma_x^{m} \, H \, R_{z}(\phi)$, again up to a Pauli transformation. Furthermore, any single-qubit rotation $U\in SU(2)$ can be decomposed as $U = R_{z}(\gamma) \, R_{x}(\beta) \, R_{z}(\alpha)$, where $\alpha$, $\beta$ and $\gamma$ are the three Euler angles of the arbitrary rotation. To be able to implement any single-qubit unitary using one-way cluster computing we require a linear 4-qubit cluster. With the same method as before the final unitary implemented on qubit \#  4 is 
\begin{equation}\label{eq:Udecomposed2}
	U= \sigma_x^{m_{3}} \, \sigma_z^{m_{2}} \, \sigma_x^{m_{1}} \, H \, R_{z}((-1)^{m_{2}}\gamma) \,  R_{x}((-1)^{m_{1}}\beta) \, R_{z}(\alpha).
\end{equation}
This form reveals most clearly how the subsequent measurements in the one-way cluster computation have to be adapted. The angles of the rotations and therefore the final correction depend on the outcomes of previous measurements, specified by the variables $m_1, m_2$ and $m_3$, which imposes a time ordering in the whole scheme. 
Equivalently to applying active single-qubit operations such as rotations and then measuring in the computational $z$-basis, one can immediately include the single-qubit operations into the measurement process and allow instead for arbitrary single-qubit measurements in the $x-y$ plane.


The above method lies at the heart of one-way computation and shows that local, single-qubit measurements are sufficient to manipulate the whole cluster. More explicitly, any computation in the one-way computer is driven by \emph{computational measurements} like the above, i.e. arbitrary single-qubit measurements in the $x-y$ plane specified by the angle $\phi$. The binary outcome of the computational measurements is always random and the result is fed forward to determine, together with the actual algorithm of the implemented computational problem, which qubit to measure next and in which measurement basis $\phi$. All single-qubit measurements that can be implemented simultaneously form an `equal-time' set $Q_t$\footnote{See \cite{RaussendorfBriegel} for a more detailed description of the computational model underlying the one-way computer.}. In general there are a number of such sets that needs to be applied in a particular successive order thereby creating a time-arrow in the computational process and we may assign one time-unit to each equal-time measurement set $Q_t$.  Manipulating the cluster state by measuring parts of it will affect the rest of the cluster through the entanglement that correlates the parts. This process is what makes the computer compute, see lower schematic in Fig.~\ref{fig:2Dspins}.


After a sequence of measurements, at time $t_{\textrm{end}}$, the remaining, so far unmeasured qubits of the cluster state are ready for readout. \emph{Readout measurements} are simple single-qubit projections in the $z$-basis and can be implemented at one instance in time altogether. The results of the measurements after $t_{\textrm{end}}$ are fully deterministic, i.e. for any given input, the readout measurements will produce a particular bit-value without ambiguity. With every deterministic readout measurement we thus find one bit of information  and the set of all of them presents the correct solution of the initial computational problem with total information $I_{\textrm{max}}$.
However, if we were to stop the computational phase prematurely, this would result in a probabilistic scenario where only part of the final output may already be deterministically available, i.e. some qubits have become disentangled from all other qubits and are either in state $|0\rangle$ or in $|1\rangle$ with respect to the $z$-basis, while other qubits are still in superpositions with respect to the $z$-basis thereby making the results of their readout measurements probabilistic. 
In the graphic on the right in Fig.~\ref{fig:magnet} we depict the rising information $I(t)$ (i.e. deterministic knowledge of bits of the outcome) with proceeding time assuming that we have already performed some equal-time measurement sets $Q_t$ up to time $t$ and then read out. There are in fact two critical times. The first critical value $t_{\textrm{crit}}$ appears, when our information about the final result becomes non-zero, i.e. when we start to gain partial knowledge of the outcome-bits by reading out. However, we have to wait until $t_{\textrm{end}}$ where the full algorithm has been implemented to obtain the complete output, $I_{\textrm{max}}$. 


We will proceed to compare one-way cluster computation to a thermodynamic phase transition process, the occurrence of magnetisation in the Ising model. Before we go into the analogy let us here first summarise the concept of a phase transition and restate Peierls' argument why there can be none in the 1D Ising model.

\section{Phase transitions and Peierls' argument} \label{sec:Peierls}

A phase transition \cite{PhaseTs} is a macroscopically observable effect which occurs when a parameter of the system is tuned across a critical value, beyond which the system takes on a qualitatively different phase. For instance, an ordered phase could be a magnetised piece of metal or  Bose-Einstein condensate (BEC) in a Bose-gas. The notion of phase transition is intimately related to that of criticality which is a concept of central importance in solid state physics. Macroscopic objects, in the form of solids, liquids and gases, undergo a diverse range of phase transitions under variation of the temperature, or external field, or pressure for instance.


When the two-dimensional Ising spin lattice undergoes a phase transition at its critical temperature, this means that above this temperature, the spins were in a randomised, disordered state, where all directions are equally likely, but below that temperature all the spins align and point in some (randomly chosen) direction, see upper schematic in Fig.~\ref{fig:2Dspins}. This transition from a disordered to an ordered state is signified by spontaneous symmetry breaking of the z-direction invariance of the state of the Ising lattice. An additional parameter has to be introduced to specify which particular `order' was chosen by the system. The \emph{order parameter} in this case is the total magnetisation, which is finite below the point of criticality and becomes suddenly zero at and above this point. Symmetry breaking is, therefore, another indicator of phase transitions.


The main advantage of the concept of criticality is the fact that wide ranging systems behave very similarly to each other at the critical point. This realisation leads to the concept of universality, namely that correlations, and consequently other observable system properties can, when critical, be described using only a small number of parameters, that are completely independent of the nature of the system.  As a result, one can derive some rather general conclusions about the existence of phase transitions.
It is well-known that there are no discrete or continuous order parameter phase transitions in 1D, no matter what particular system we are talking about, providing that the interactions are short ranged. This result has been proven using many different methods, first by Peierls (and tidied-up by Griffiths) \cite{Peierls}, then by Mermin and Wagner \cite{MerminWagner}, and  finally by Hohenberg \cite{Hohenberg}. The gist of all these arguments is that any thermal fluctuations in 1D are enough to destroy order that is generated by short range interactions. In other words, entropy always dominates energy in 1D. In this paper we will rely on the argument due to Peierls since it is the most suitable one for our application to one-way computation.

\bigskip

The intuitive picture of Peierls' argument \cite{Peierls} is as follows. The classical one-dimensional spin chain interacting with the Ising short range coupling would undergo a phase transition from a disordered phase to an ordered phase if a non-zero magnetisation arose below a certain critical temperature. To test whether the ordered state can be an equilibrium state of the system for any finite temperature, we assume that the system is initially ordered and then perturb the state slightly while observing the change of the free energy,  $\Delta F$, with 
\begin{equation} \label{eq:freeenergy}
    F = U - T \, S,
\end{equation}
where $U$ is the internal energy, $T$ the temperature and $S$ the entropy of the state. If the ordered state was an equilibrium state, then its free energy must be minimal and any perturbation out of this state would increase free energy, $\Delta F > 0$.  If, on the other hand, there exists a perturbed configuration that can decrease the free energy, the system will tend towards this new configuration preferring it energetically over the ordered state. This immediately implies that any order will be destroyed and no criticality is exhibited.


The Hamiltonian of the 1D Ising spin chain with nearest neighbour interaction is given by
\begin{equation}
    H = -  J \sum_{j=1}^N \sigma_j \sigma_{j+1},
\end{equation}
where the $\sigma_j$ are the Pauli spin operators in $z$-direction for the $j$-th qubit  and $N$ is the number of qubits in the chain. The interaction $J$ is positive in the ferromagnetic case and negative in the anti-ferromagnetic case. The ordered state is reached when all spins are aligned, either in the positive or negative $z$-direction. We proceed by applying a single random flip, which realises the smallest possible perturbation. The change of internal energy under such perturbation is $\Delta U = \langle H \rangle_{\textrm{flipped}} -  \langle H \rangle_{\textrm{ordered}}  = -J (N-2) - (-J N) = 2 J$, because every spin is connected to two nearest neighbours. There are $N$ different ways to flip a single spin so that the change in entropy is $\Delta S = k_B \ln N$. Therefore the change in free energy is
 \begin{equation}
    \Delta F = 2 J - k_B T  \ln N,
\end{equation}
and since we are looking at the thermodynamical limit of large $N$ it is clear that $\Delta F$ will always be negative for any finite temperature. The conclusion is thus, that in 1D no ordered state can exist in a finite temperature equilibrium. The entropy, i.e. the thermal fluctuations, will always outweigh the tendency of energy to create order. Similar conclusions can be drawn for all short range interactions completing Peierls' proof that no phase transition takes place in one dimension.

\begin{figure}[t]
    \begin{center}
    \includegraphics[width=3cm]{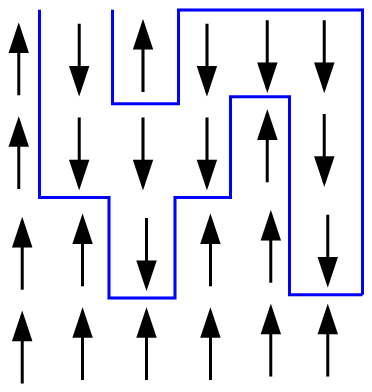}
        \caption{\label{fig:2Dcut} How to flip spins in a 2D lattice in the best case scenario. This figure shows a possible flipping pattern that cuts the lattice into domains of equal alignment. A path that wiggles back and forth inverts a lot of nearest neighbour couplings and the energy of the lattice increases considerably (maximally). However there are many similar, albeit different such paths so that the entropy of the lattice increases at the same rate as the energy. Therefore, there exists a finite temperature range, where the entropic fluctuations are more costly than the energy gain and the fully aligned (ordered) configuration remains the minimum of free energy. Above this critical temperature, the ordered phase is destroyed and a phase transition occurs.}
    \end{center}
\end{figure}


This situation is changed dramatically in two and higher dimensions. In two dimensions the ordered state is also the one where all the spins point in the same direction. However, flipping a single spin is now no longer enough to destroy this order; instead the smallest perturbation must cut the lattice into two different domains\footnote{If we were to flip only single spins here and there the two-dimensional lattice would still be connected as one phase and in the thermodynamic limit these errors wouldn't matter. To properly bound or separate a two-dimensional object one needs a one-dimensional line.}.
Let us now calculate the best case scenario for such a cut to destroy order. As far as the balance between change in energy and entropy is concerned it is the one which cuts the lattice into two halves by crossing as many nearest neighbour couplings as possible along the way. For such a cut, the change in energy is at most $2 N J$, since this is the number of couplings present in the lattice (see Fig. \ref{fig:2Dcut} for an illustration). The entropy change, on the other hand, is roughly $k_B\ln 3^N$ because after each spin we have a choice to proceed with the cut in at most three possible directions (everywhere but backwards). The change in free energy for this kind of perturbation is thus
\begin{equation}
    \Delta F = 2NJ - k_B T N \ln 3.
\end{equation}
By setting $\Delta F =0$ we obtain the following critical temperature for the stability of the ordered phase
\begin{equation}
    T_{\textrm{crit}} = {2 J \over k_B \ln 3}.
\end{equation}
This is astonishingly close to the exact temperature obtained by Onsager \cite{Onsager} ($ T = 2.27 {J \over k_B}$)  especially given that the above argument is very simplistic. 


To summarise, in two dimensions the two competing quantities can be balanced and a phase transition can occur at a finite temperature. With the same argument one can show that this holds also true for higher dimensions. We will now proceed to draw the analogy between the phase transition in the Ising model and the process of one-way cluster computing.

\section{The computational analogy} \label{sec:analogy}

A central question for one-way quantum computation is, `Why does it work?' Although it is possible to go through each mathematical step, using the rules of one-way computing, it is not obvious \emph{how} the computational power arises. Clearly, the existing correlations in the cluster state have to be traded somehow for the information about the solution of the computational problem. But how can we find the  optimal way to do it? And what is the ingredient that makes, apparently, one-way quantum computers more powerful than their classical counterparts?

\begin{figure}[h]
    \begin{center}
    \hspace{-2cm}\includegraphics[width=9cm]{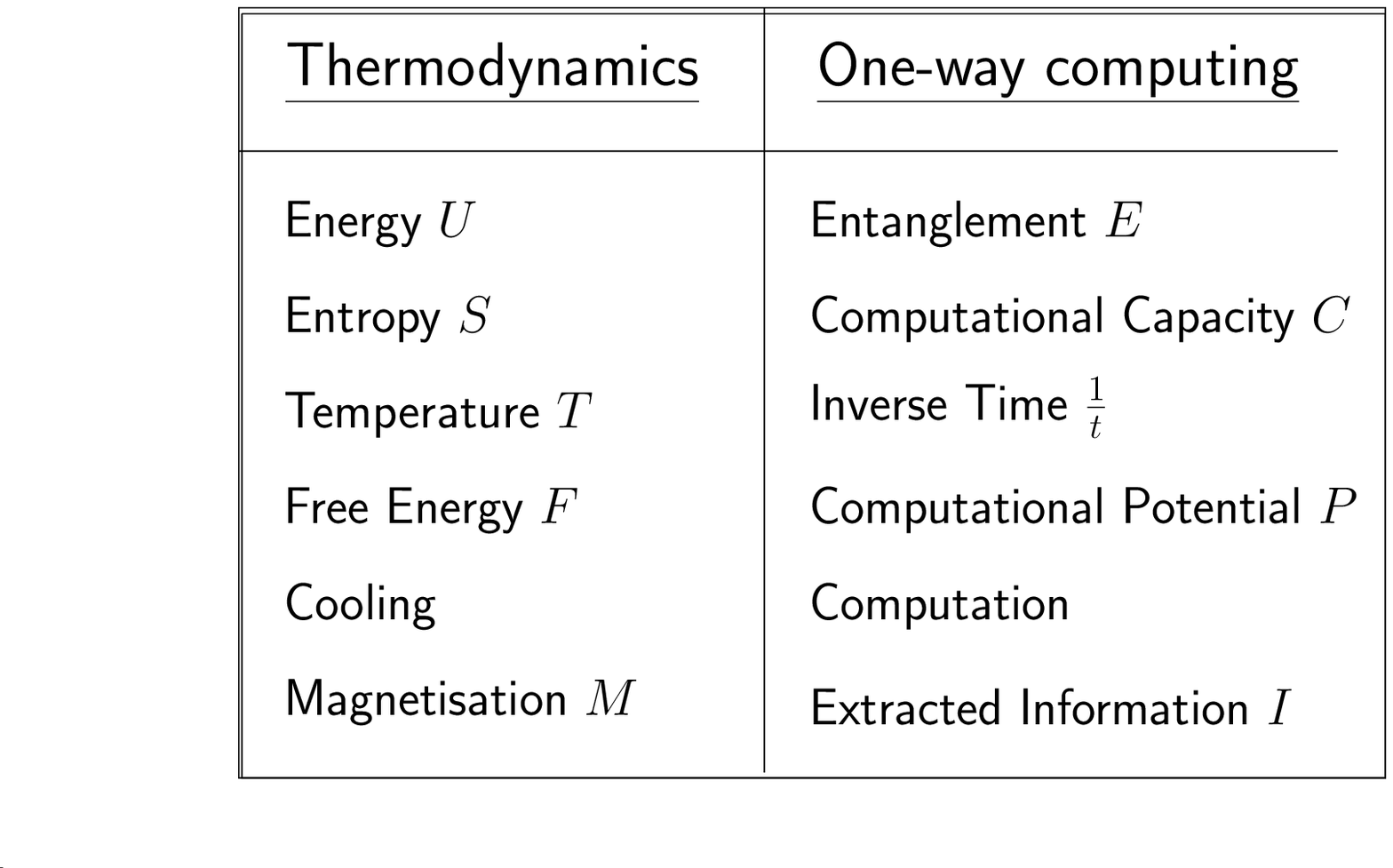}
    \caption{\label{fig:comparison} This table compares the main quantities relevant in a thermodynamical phase transition process and in the one-way computer model. For details please refer to the text.}
    \end{center}
\end{figure}

In the following we compare the process of one-way computation to the phase transition in the Ising model and identify potentially useful quantities for the one-way computer in analogy to those used in thermodynamics, as shown in the table in Fig.~\ref{fig:comparison}. Our comparison cannot directly answer any detailed questions yet. However, the ideas sketched here open up a completely new way of looking at the dynamics of the one-way computer. Using the language of phase transitions it becomes intuitive why one-way computing can work and  it also sheds light on how the computational process can be traced with only a few characteristic variables. The analogy will furthermore allow us to give rough estimates of minimal requirements for universal quantum computing.
One result we can re-produce is to heuristically understand why the 1D cluster state is not a universal resource for one-way computation whereas the 2D cluster state has this potential. Additionally, we will present an estimate of the critical time for a computation depending on the dimension or connectivity of the resource state for dimensions higher then 1D. 


We saw how cooling a magnetic material of a multitude of spins means that the individual spins are flipped to obtain a new thermal equilibrium between two quantities, the energy and the entropy. Thermal equilibrium is determined by the second law of thermodynamics as the configuration that maximises the entropy for fixed internal energy or, to put it differently, the configuration that achieves the lowest possible free energy for a given temperature. We propose that an optimal one-way computer should work in principle in the same way as nature. Similarly to thermal systems, we expect a principle of least action that balances two quantities at each step in the computation to maintain enough computational power to finalise the computation. These two quantities are the \emph{entanglement} and the \emph{computational capacity} which will be defined in the following.

\begin{figure*}[t]
    \begin{center}
    {\includegraphics[width=10cm]{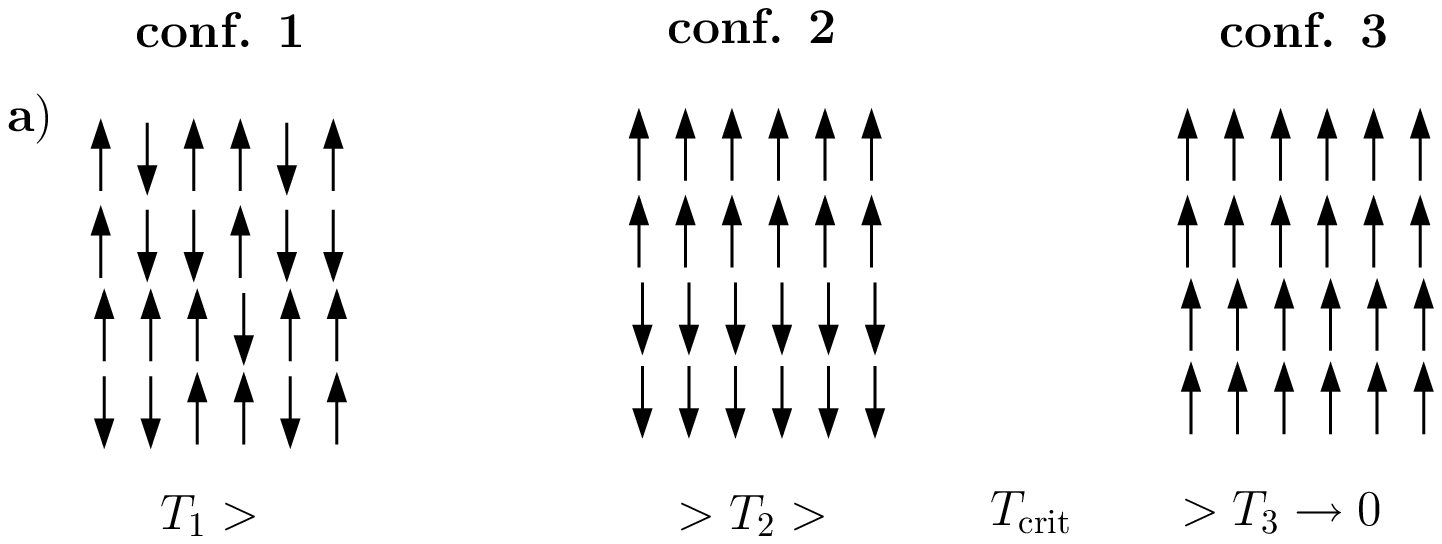}}\\[2ex]
    {\includegraphics[width=10cm]{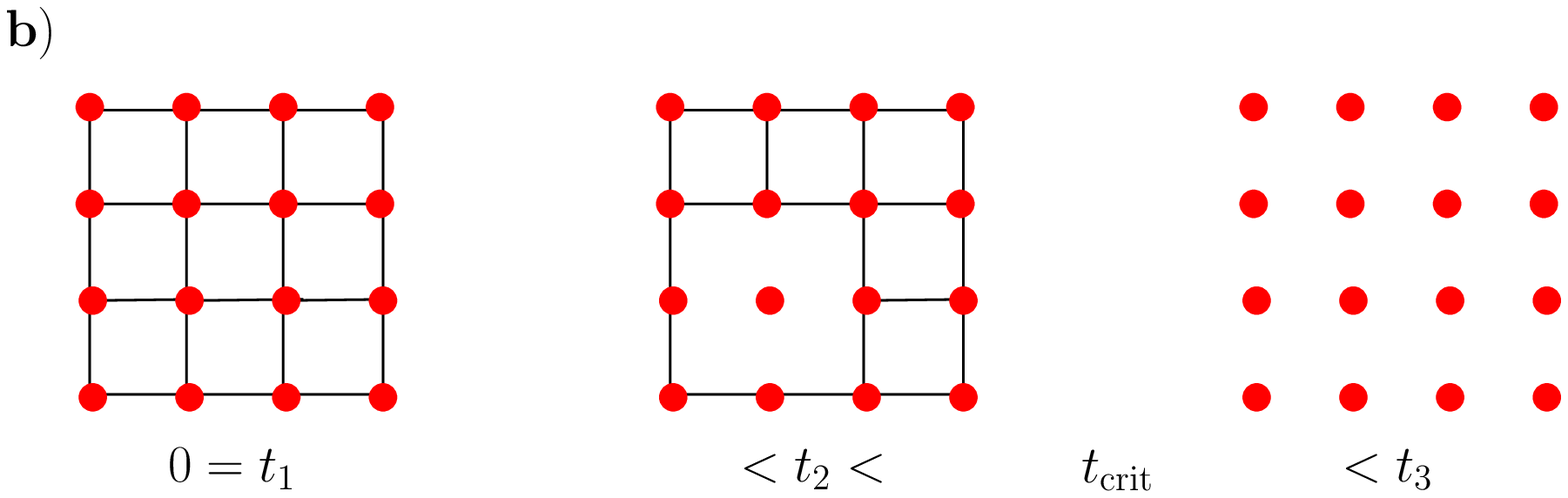}}
        \caption{\label{fig:2Dspins} Graphical comparison of the Ising model and the one-way computer. The upper graphic, \textbf{a)}, shows the formation of an ordered (magnetic) phase of spins with decreasing temperature in the 2D Ising model. The lower graphic, \textbf{b)}, shows the evolution of the 2D cluster computer with increasing time to the ordered state of completely disentangled qubits. When this state is reached a particular solution of the computation has been chosen and can be read out from the qubits by simple measurements. The different stages during these processes are shortly summarised here. 
        \textbf{a)}
	\textbf{conf. 1: } $T_1 \sim$ big: Spins are randomly flipped so that half of them point up and half of them down, without any kind of order. The magnetisation is zero, $M_1= 0$. 
	\textbf{conf. 2: } $T_2 \gtrsim T_{\textrm{crit}}$: Spins in the lower half and upper half point in opposite directions and form a Weiss-domain each with distinct magnetisation. Nevertheless, the total magnetisation is still zero.  
	\textbf{conf. 3: } $T_3 \to 0$: All spins point in one direction and form one ordered phase with $M_3 = M_{\max}$. 
	The critical temperature, $T_{\textrm{crit}}$, for this phase transition is passed somewhere in between \textbf{conf. 2} and \textbf{conf. 3}. 
	\textbf{b)}
	 \textbf{conf. 1: } $t_1 = 0$: Start of computation with full cluster state. All qubits are equally entangled and no bias towards any solution is taken, $I_1 =0$.
	 \textbf{conf. 2: } $t_2 > 0$: A few measurements have been made, some qubits are therefore not entangled with the cluster anymore, however no useful information has become available yet. 
	 \textbf{conf. 3: } $t_3 \to t_{\textrm{end}}$: All qubits have becomes disentangled and ordered and their value can be read out. The solution of the computation is thus fully available and the information is $I_3 = I_{\max}$.
	 The critical time, $t_{\textrm{crit}}$, at which we start to gain knowledge about the solution of the computational problem is passed somewhere in between \textbf{conf. 2} and \textbf{conf. 3}.  
 }
    \end{center}
\end{figure*}


While in the Ising model each spin interacts via Heisenberg coupling with its nearest neighbour, in the one-way cluster computer each pair of nearest neighbouring qubits is C-phased, see Fig.~\ref{fig:2Dspins}. Even more, this C-phase gate operation can be achieved by letting precisely the Heisenberg interaction act on the qubits for a certain amount of time. In the Ising model, the Heisenberg interaction results in a non-trivial internal energy $U$. In thermodynamics the internal energy is the resource for a physical system to perform work and, in the context of phase transitions, energy provides the capability to create order. It is appealing and intuitive to view entanglement in much the same way for the one-way computer. The entanglement, $E$,  which is generated by the C-phasing, here allows the transport of information across the cluster state, eventually resulting in the ordering of the possible outcomes of a computational task into a single, unique solution. It is thus the resource for executing dynamics in the one-way computer. Each spin flip in the Ising model costs one unit of energy, similarly each computational measurement in the one-way computer consumes approximately one unit of entanglement. The number of gates implemented in this way, i.e. the computational `work', is proportional to the number of units of entanglement used up by the measurements\footnote{Linking the use of entanglement to the production of computational work has been proposed before, e.g. in \cite{Horodecki} it was argued that teleporting qubits is analogous to work in the context of state distillation. For one-way computation this intuition now carries over to general quantum state manipulations. Other analogies between the distribution of entanglement and thermodynamical models have been exploited. For example, entanglement distillation has been successfully linked to thermodynamical heat engines \cite{16,17} resulting in a better understanding of why the reduced von Neumann entropy quantifies the amount of entanglement in pure states}.


On the other hand, the entropy, $S$, counts the (logarithm of the) number of microscopic configurations or the multiplicity of a configuration. In the Ising model this is the number of configurations of spins that lead to the \emph{same} energy. The change of entropy from a state with energy $U_1$ to another with energy $U_2$ is the number of ways in which spins can be flipped to reach the new state, i.e. the number of ways the same energy difference, $U_2 - U_1$, can be achieved. Let us define an analogous quantity for one-way computation, the \emph{computational capacity}, $C$, that counts the number of configurations of qubits with the same amount of entanglement, $E$. The change of computational capacity, $\Delta C$, is again the number of ways in which the same difference in entanglement can be achieved. Due to the non-reversible nature of the one-way computer, a direction in time is set and any state with entanglement $E_1$ will always evolve to a state with less entanglement, $E_2 < E_1$. The change of computational capacity $\Delta C = C_2 - C_1$ when measuring and transforming a state with $E_1$ into a state with $E_2$ is the number of possible evolutions (c.f. Toffoli's definition in \cite{Toffoli}) that lead from $E_1$ to $E_2$. The number of different evolutions of the computer is just the number of different computations, hence  $\Delta C$ also expresses the capacity the initial state has to compute for a given reduction of entanglement. In analogy with thermodynamics and to allow for an additivity law for combining systems similar to the ones in statistical mechanics we define the computational capacity to be the  \emph{logarithm} of the number of ways in which the entanglement can be reduced. 


We note however, though the amount of contained entanglement can be calculated in symmetric multi-qubit states \cite{Markham}, the evaluation becomes very difficult for non-symmetric states. Thus, it is not clear how to specifically calculate the computational capacity for a partially measured cluster state. Yet the hope is, that at least bounds on the entanglement can be derived that are based on the symmetric portion of the remaining cluster. 
Having established the similarities of energy and entanglement, and entropy and computational capacity, what are the equivalents of the other parameters that feature in a phase transition? Clearly, when cooling the Ising model the temperature is the process parameter that is continuously  reduced. How does the temperature come in and what is its analogue in the one-way computer?


The set of states of a thermal system that balance the energy and entropy to minimise the free energy can be described by a curve in the set of all states, parametrised by a single parameter, the temperature, $T$. These equilibrium states are the thermal states. In the Ising model, the temperature can also be understood as setting the scale of how costly a perturbation from the ordered state is in terms of entropy. In one-way  computation the state of the computer is changed by \emph{us} as we measure the cluster to drive a certain computation forward. The produced curve of states that is generated by these successive manipulations is again a curve in state space and can be parameterised. Indeed, since the time $t$ moves forward monotonously during the computation we choose the time as the parameter of the curve or, for reasons that will become apparent later (c.f. Fig.~\ref{fig:2Dspins}), we choose ${1 \over t}$. The inverse time, ${1 \over t}$, tells us how much each computational step costs. Yet, $t$ is not simply proportional to the size ($N$) of the cluster state since one can perform more than one measurement in a single computational step. In fact, all measurements belonging to an equal-time set $Q_t$ can be done in one instance of time (by definition of the sets $Q_t$). Thus, $t$ defines a measure of the depth of the computation and will be used to count the number of equal-time sets needed to successfully obtain the  solution to a computation.


Finally, the central role in a phase transition process is played by the free energy which must be balanced across the phases. In general thermodynamics, the free energy determines how much of the total energy, $U = F + T \, S$, is available to perform work and what portion,  $T \, S$, needs to remain to maintain thermal fluctuations at a given temperature. A state of thermal equilibrium is reached when the free energy is minimal. To achieve the minimal free energy a system may take on qualitatively different configurations, the phases, which alter the use of the available energy. This qualitative change of configuration is what we call a phase transition and the temperature at  which this happens is the critical temperature. For the one-way computer we define the \emph{computational potential}
\begin{equation}\label{eq:comppotential}
P \sim E - {1 \over t} \, C,
\end{equation} 
as the quantity that balances the available  entanglement with the computational capacity of the remaining resource state at each time step in the computation, much like the free energy in Eq. \eqref{eq:freeenergy} balances the energy and the entropy for each temperature. We propose that the best way to run the one-way computer is to always minimise the computational potential during computation. Apart from the analogy to the phase transition process the reason is the following.


Universality in the context of the one-way cluster computer means that for any computational task there exists a cluster state of $N$ qubits, with $N$ big enough to produce the solution to the computational task. Measuring a cluster state in such a way that the largest number of evolutions (or computations) remains, leaves the cluster computer most `universal'. So, any computational process should be run in such a way, that the number of possible evolutions, the computational capacity $C$, of the remaining resource state, at each step in the computation, is maximised for a given reduction of the entanglement, $E$. Maximising $C$ for a given $E$ and time corresponds exactly to a minimisation of the computational potential $P$. This is analogous to thermodynamics where the maximisation of the entropy, i.e. the maximisation of choices, for a given internal energy leads to the minimisation of the free energy. Of course, computations could be run in a less resourceful way, but for our argument it is sufficient to discuss the optimal strategy. This is our proposed law of least action for the dynamics of the one-way computer. In analogy to the free energy, the computational potential thus measures how much of the entanglement in the resource state is `free' to be used for the computation while keeping enough structure of connecting entanglement. 


\begin{figure*}[t]
    \begin{center}
    {\includegraphics[width=11cm]{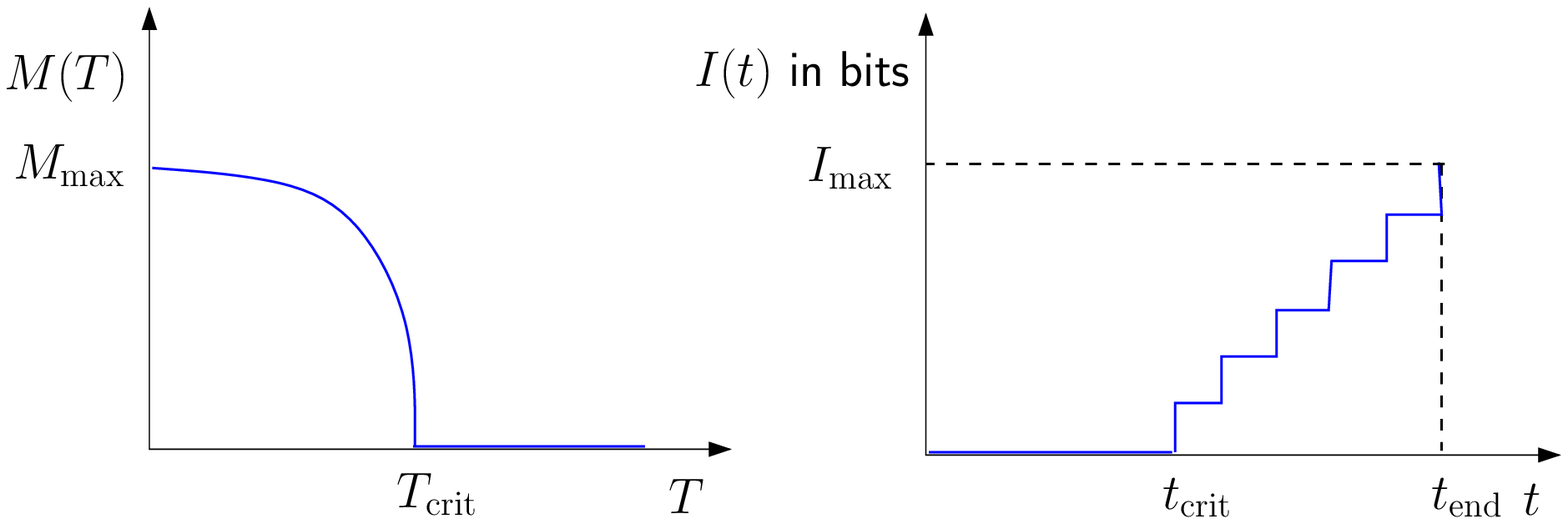} }
        \caption{\label{fig:magnet}  Order parameters in the 2D Ising model (magnetisation $M (T)$) and the 2D cluster computer (information $I (t)$) are sketched over their process parameters, temperature $T$ and time $t$, respectively (they are inversely proportional as indicated in the text). In 2D there exists a finite $T$ ($t$) at which the order parameter $M$ ($I$) becomes non-zero.  In the cluster state computation in the first stage ($t < t_{\textrm{crit}}$) all measurements are performed in the equatorial $x-y$-plane to drive the computation. The measurement outcomes have always random outputs, i.e.  the probability for either of the two outcomes is always $1/2$.  
Outputs remain random until the computational phase reaches $t_{\textrm{crit}}$, at which the first qubits become disentangled from all others and their value is deterministically available. The actual outcomes are then worth one bit of information for each measured qubit. To obtain the full solution however, the full algorithm has to be implemented which is finished only at $t_{\textrm{end}}$. The total available information  grows with time to the maximal information of the outcome $I_{\max}$ at $t_{\textrm{end}}$. }
    \end{center}
\end{figure*}

We have identified general thermodynamic quantities, such as the internal energy, the entropy, the temperature and the free energy, that feature in all thermal processes. We have also established a law analogous to the second law of thermodynamics that makes use of the resource state in the optimal manner. Nevertheless, in a phase transition processes an additional parameter is needed to describe the system fully. This \emph{order-parameter} appears when a new, ordered phase emerges, for instance the magnetic phase in the Ising model. The parameter in this example is the  magnetisation $M$ and it specifies in which spatial direction the magnetisation points. In other words it serves to specify how the magnetic material has broken the symmetry of the governing Hamiltonian. In one-way computation, `symmetry' is broken as soon as we start to learn something about the solution to the computation. Different solutions of the computation are like the different choices of the directions for the magnetisation. The order-parameter for one-way computing, analogous to the amount of magnetisation in the Ising model, is then proportional to the number of retrieved bits of information of the solution, $I$. Since the measurements in the first stage of one-way computation are just for preparation purposes, they do not give us any information. This implies that no `choice' has yet been made and the information about the solution remains zero, $I(t) =0$ for $t < t_{\textrm{crit}}$ (early times $\sim$ high temperatures). However, at some time $t_{\textrm{crit}}$ a part of the former cluster state may already be deterministically available for readout, i.e. some qubits have become disentangled from all other qubits and are either in state $|0\rangle$ or in $|1\rangle$ while other qubits are still in superpositions making the results of their readout measurements probabilistic. Finally, when the proper read-out stage is reached at $t_{\textrm{end}}$ the full outcome is available and the information becomes maximal, $I(t_{\textrm{end}}) = I_{\max}$. The growths of information over increasing time in the computation thus resembles the rise of magnetisation with decreasing temperature in the Ising model and the two curves are displayed schematically in Fig.~\ref{fig:magnet}.

\bigskip

\textbf{Peierls' argument for the cluster computer}
Having identified the computational characteristics that correspond to the thermodynamic quantities featuring in a phase transition process, we now return to Peierls' argument and apply it to one-way computation. Let us regard the final state of the computation, where all qubits have been measured and all entanglement has vanished as the \emph{ordered state of the computation}. Actually, what has become ordered is the \emph{deterministic state} of each single qubit in the cluster that implies the full knowledge of the solution of the computation. We can now ask the same question as in the case of the Ising model: Is the ordered state stable, or in the language of computation, will the solution ever be reached and the computation comes to an end? Again, consider first the 1D case of a chain of $N$ qubits. The final state of the cluster computer can be depicted schematically as below.
	\begin{center} 
		\textbf{final state} \,\,\,\,\,\,\, {\includegraphics[width=2.5cm]{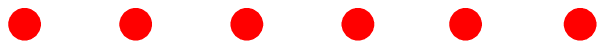}} 
	\end{center}
The smallest perturbation that disturbs the order is a single pair of entangled qubits, which leaves the last bit of information unavailable, i.e. the phase of completely disentangled, ready to be read-out qubits is interrupted.
	\begin{center} 
		\textbf{origins} \, {\includegraphics[width=2.5cm]{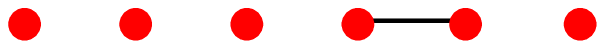}},
    	      		\, {\includegraphics[width=2.5cm]{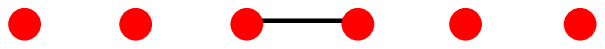}},
    	     		 \, {\includegraphics[width=2.5cm]{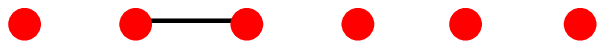}}, ... 
	\end{center}
All states of this kind form the set of possible origins before reaching the final, ordered state. Again, like in Peierls' argument we find $N$ different evolutions towards the ordered state. This implies an increase of the computational capacity under the perturbation by $\Delta C = \ln N$. The change of entanglement on the other hand  is just one unit of entanglement, $\Delta E =1$, and hence the change of the computational potential is $\Delta P = 1 - {1 \over t} \ln N$. The change should be positive for some time $t$ to allow the final step of the computation to actually be executed at that time. However, the number of qubits in the cluster, $N$, can be arbitrarily large to allow more and more complex computations. For these increasingly bigger cluster states the change of the computational potential is never negative, unless the time is infinitely large ($t \propto \ln N \to \infty$).  In other words, more complex computations will not reach the final solution in a finite time and the one-dimensional cluster states are therefore not universal. 
   

But what is the reason that the final state cannot be reached? Let us take a look at the beginning of the computation to get an idea of what goes wrong.  Initially we have the fully symmetric cluster state and we start the computation by measuring at least one qubit. The number of ways in which this can be done is $N$ again and the change of the computational capacity is $\ln N$ while we have reduced the entanglement by only 2 units. Physically this means that we have lost far too many possible computations ($\ln N$) compared to the number of computational gates that have been executed. While this imbalance may still permit us to solve a limited number of computations, it is not enough to achieve universal computation, even when we let the size of the initial cluster state grow larger and larger. 
To summarise, what is important for one-way computers is the trade-off between the entanglement that they contain and the computational capacity. The former represents the actual gates that our computer is able to execute, while the latter is the number of all possible computations that we can in principle perform. In 1D, the change in computational capacity by far outweighs the number of gates implemented and this is why we cannot perform any computation we desire. The cluster computer gets stuck at some point. 
Following the same argument one can show that in two dimensions universality is at least possible because the two terms in the computational potential grow at the same rate. Indeed, it is known that 2D clusters are universal from a straightforward constructive argument \cite{RaussendorfBriegel}. In the last section we will now quantify how the two quantities can be balanced in higher dimensions and with longer ranged correlations. 

\section{Implications and Conclusions} \label{sec:conclusions}

A number of implications are generated by the conjectured analogy. Here we choose to focus on two main ones. First, we look at how the efficiency of computation scales with the dimensionality of the computer. Secondly, we look at the effect of allowing long range C-phases to generate entanglement between distant qubits while still remaining in 1D.


Let us calculate the change in the computational potential for a computer in $d$ spatial  dimensions. Then each qubit has at most $2d$ nearest neighbours. Note, that a low-dimensional system can also effectively become higher dimensional by increasing the interconnectivity between the qubits via non-nearest neighbour C-phase gates. We now repeat the argument of Peierls' for general $d$ dimensions. After little inspection, it is clear that the optimal Peierls' cut reduces entanglement by an amount proportional to $d \, N$. At the same time, the number of possible possible cuts to achieve the same configuration is proportional to $\ln d^N$. Balancing the two, to maximise the computational potential we obtain 
\begin{equation}
    {1 \over t} \ln d^N \propto d N,
\end{equation}
giving us the critical time for computation of the form
\begin{equation}
    t_{\textrm{crit}}  \propto {\ln d \over d}.
\end{equation}
This prediction agrees with our intuition that computation should be faster with increased dimensionality because there is more connectivity inside the cluster in higher dimensions.


Secondly we turn to the issue of how to make a one-dimensional cluster computer universal. It is clear that we have to increase the interconnectivity between the qubits in some way. We can do that by allowing C-phase gates to act not only between nearest neighbours, but also between more distant qubits, i.e. the C-phase gates become long range. Therefore Peierls' argument has to be reviewed for the case of long range interactions. This was done in 1969 by Thouless in \cite{Thouless} and the result, within our context, is that if C-phases drop faster than the square of the distance between the qubits, then we cannot achieve universality of computing. A slower drop with distance could possibly be sufficient to allow universal computation. Any successful one-way model architecture will have to take this issue of connectivity into account.

\bigskip

In this paper we have argued that one-way quantum computation can be viewed and understood as a phase transition.  In standard phase transitions temperature is the parameter which determines which phase the system is in. In computing, this role is played by time. However, it is transparent that temperature also has a role to play in computation. After all, should the ideal cluster computer be exposed to an environment of non-zero temperature, the thermal fluctuations would destroy the perfectly symmetric arrangement of exact C-phase gates. The higher the temperature of the environment, the more mixed the states of the cluster computer will be. At some high enough temperature this mixing will destroy so much of the available entanglement that the computer will lose all its quantum computational power. (For a discussion of the effect of thermal fluctuations in cluster computers, see \cite{1Dcase}.)


Is it possible for us to estimate this temperature beyond which the one-way quantum computer becomes classical? We know that in many-body systems entanglement exists only below some critical temperatures. A very general rule of thumb is that the temperatures should be lower than the strength of coupling between the individual systems for the whole systems to be entangled \cite{Janet2}.  An interesting conclusion can now be drawn when this result is translated to cluster computing. Can we say that the cluster state is only able to be a universal quantum computer at temperatures smaller than the coupling constant for generating the C-phases?


We leave this, along with many other issues, as an open question. Our conclusion is that the analogy between thermodynamics and computation allows the flow of ideas between two established theories and promises mutual benefit. While some of the raised questions may be difficult to resolve, we hope that our analogy will stimulate further research into this exciting area.

\begin{acknowledgements}
The authors thank A. Miyake for insightful discussions. J. A. is supported by the Gottlieb Daimler und Karl Benz-Stiftung. M. H. \& V. V. would like to thank the Engineering and Physical Sciences Research Council in UK and the European Union for financial support. This work was supported in part by the Singapore A*STAR Temasek Grant. No. 012-104-0040. D. M. acknowledges support of JSPS. J. A. acknowledges funding by the QIP IRC network. V.V. acknowledges financial support from the Royal Society and the Wolfson Trust in the UK as well as the National Research Foundation and Ministry of Education in Singapore.
\end{acknowledgements}


\end{document}